\documentclass[runningheads,a4paper]{llncs}

\usepackage{amssymb}
\usepackage{amsmath}
\usepackage{graphicx}
\usepackage{booktabs}
\usepackage{tikz}
\usetikzlibrary{arrows.meta,positioning,calc}
\usepackage{url}
\providecommand{\keywords}[1]{\def\and{{\textperiodcentered} }%
\par\addvspace\baselineskip
\noindent\keywordname\enspace\ignorespaces#1}

\newcommand{\PIV}{\mathrm{PIV}}
\newcommand{\Pdec}{P_{\mathrm{dec}}}
\newcommand{\sd}{\mathrm{sp}}

\begin{document}

\mainmatter

\title{Physiological Information Reliability: Contextual-Bandit Allocation of Coding Redundancy and Computation in Multimodal Cardiovascular Sensing}
\titlerunning{Physiological Information Reliability}

\author{Navaneetha Krishnan Kamalakannan\inst{1}
\and Janakiraman Kamalakannan\inst{2}
\and Harinisri Velmurugan\inst{3}}
\authorrunning{N. K. Kamalakannan et al.}

\institute{Department of Electronics and Communication Engineering,\\
Saveetha School of Engineering, Chennai, India\\
\email{knavaneeth385@gmail.com}
\and
Government Kilpauk Medical College and Hospital, Chennai, India\\
\email{kjanaki2408@gmail.com}
\and
Department of Physics and Nanotechnology,\\
SRM Institute of Science and Technology, Chennai, India\\
\email{harinisriv2003@gmail.com}}

\toctitle{Physiological Information Reliability}
\tocauthor{N. K. Kamalakannan, J. Kamalakannan, H. Velmurugan}
\maketitle

\begin{abstract}
A packet can arrive intact and still be nearly worthless. On a body-worn node the radio and processor share one small battery, yet ECG and photoplethysmography (PPG) windows differ widely in how much an estimator can recover from them. We propose Physiological Information Reliability (PIR): a hand-designed window-value proxy, PIV, built from signal quality and cross-modal heart-rate agreement, joins channel, energy and compute telemetry in one state vector, and a LinUCB bandit chooses coding redundancy and processing site. Over 25 paired seeds on a synthetic burst-erasure testbed, PIR-LinUCB used 4.3\% less energy than a hand-written rule, with lower error and higher delivery. A rule that also reads PIV matches it on error and delivery but spends 3.7~mJ more per window. On five real fetal ECG recordings, PIV tracked heart-rate error better than signal quality alone (mean within-record Spearman $\rho=-0.50$ against $-0.40$). The evidence is simulated, small and non-clinical.
\end{abstract}

\section{Introduction}

A sensor does not know whether its output is any good. It hands over a window of samples, and the radio, the scheduler and the estimator downstream treat that window as one more unit of payload. Windows differ, though. Motion drives an optical channel into noise, an electrode lifts, one modality drops out while the other carries on. That bits arrived and that they are worth anything are separate questions, usually answered in separate layers. The gap costs most on a body-worn node, which pays for every radio packet (2~mJ in our model), shares a bursty channel, and must decide window by window how much redundancy to buy and where to run inference. Equal protection for a clean window and a corrupted one wastes energy, and too little for a clean one loses something a packet delivery ratio never shows.

The ingredients exist. Quality indices for ECG and PPG were derived with wireless monitoring in mind~\cite{orphanidou2015}. Multimodal networks reweight modalities by estimated reliability~\cite{zhang2023} or tolerate a missing one~\cite{ma2022}, but they do so inside the model; Liang et al.~\cite{liang2024} survey the wider field. Communication theory argues for preserving task-relevant content instead of bits~\cite{gunduz2023,shao2022} and for metrics that look past delivery, such as age of information~\cite{kosta2017}. As far as we could tell, these pieces are tuned separately, and none of them lets a window-level reliability estimate set the redundancy that the window receives. Our value proxy, PIV, is a crude, hand-designed relative of such measures. It is a cheap scalar available before transmission, used to decide how much to spend rather than how to weight features. We place it in one state vector with channel, battery and compute state, and let one online learner price them against each other.

The question is deliberately narrow: how should a resource-constrained multimodal sensing pipeline allocate coding redundancy and computation placement when window quality varies and the channel is bursty? We instantiate it with ECG and PPG, whose quality is easy to degrade in a controlled way and whose downstream task, heart rate, has an unambiguous reference in synthetic data. The learner is a linear contextual bandit on purpose. It has six arms and a ten-dimensional context, updates in closed form with rank-one steps, and fits the memory and arithmetic budget of the node it is meant to run on; neural bandits~\cite{zhou2020} are the obvious next step once the problem is better characterized.

\textbf{Contributions.} (i)~PIR, a joint state $X_t$ that places a window-value proxy beside wireless, energy and compute state (Sect.~\ref{sec:form}). (ii)~A LinUCB instantiation choosing redundancy $R\in\{2,8,20\}$ and local or edge placement per window. (iii)~A 25-seed comparison with five non-bandit policies, among them a hand-written rule that is given PIV, with paired inference reporting estimation error, energy, delivery and reward together under a multiplicity correction and checked with a rank test (Sect.~\ref{sec:res}). (iv)~A decomposition of the controller's advantage: PIV accounts for the gain in accuracy and delivery, learning for the energy saving. An ablation that replaces PIV in the context by a fixed reference value supports the first part. (v)~A first check of PIV on real physiology: five abdominal fetal ECG recordings, where PIV is compared against heart-rate error measured with an independent direct-electrode reference (Sect.~\ref{sec:real}).

\section{Related Work}

\textbf{Signal quality and reliability.} Signal-quality indices for ECG and PPG are an established line of work; Orphanidou et al.~\cite{orphanidou2015} derived indices for both signals with ambulatory wireless monitoring as the target use. PIV borrows the quality term and adds a penalty for disagreement between channels; it is a trust proxy, not a diagnostic quantity.

\textbf{Reliability inside multimodal models.} Dynamic fusion weights each modality by an estimate of its reliability~\cite{zhang2023}, and the sensitivity of multimodal transformers to a missing modality has been measured and partly repaired~\cite{ma2022}. Both act after the data have arrived; PIR acts earlier, on the protection and processor a window receives, and never touches the estimator.

\textbf{Task-oriented communication and offloading.} Task-oriented and semantic communication changes what is sent~\cite{gunduz2023}; edge inference with learned compression, for instance, trades rate against task accuracy through an information bottleneck~\cite{shao2022}. Age of information measures staleness, not content~\cite{kosta2017}. We keep the payload fixed and vary protection, using random linear network coding~\cite{ho2006} over a Gilbert--Elliott burst channel~\cite{gilbert1960,elliott1963}, and we let a content proxy decide how much protection a window earns. Our earlier work applied hierarchical random linear coding to decentralized wireless groups~\cite{kamalakannan2026apc}.

\textbf{Contextual bandits.} LinUCB~\cite{li2010} keeps one ridge regression per arm and adds an upper-confidence bonus; its regret was analyzed in~\cite{chu2011} and tightened in~\cite{abbasi2011}. Thompson sampling is the Bayesian alternative~\cite{agrawal2013} and NeuralUCB~\cite{zhou2020} swaps the linear model for a network; we use LinUCB because the decision must be cheap at the sensor. We found no earlier work that conditions coding redundancy on a window-level physiological reliability score jointly with compute placement, though we ran no systematic survey.

\section{The PIR Formulation}
\label{sec:form}

\begin{figure}[tbp]
\centering
\begin{tikzpicture}[
  node distance=5mm and 6mm,
  box/.style={draw, rounded corners=1.5pt, align=center, font=\scriptsize, inner sep=3pt, minimum height=9mm, text width=31mm},
  arr/.style={-{Stealth[length=1.8mm]}, line width=0.5pt}]
  \node[box] (w) {ECG and PPG window};
  \node[box, right=of w] (q) {SQI, cross-modal HR agreement $\rightarrow\PIV_t$};
  \node[box, right=of q] (x) {state $X_t=[S_m,S_w,S_e,S_c]$};
  \node[box, below=7mm of x] (c) {LinUCB controller};
  \node[box, left=of c] (a) {action $a_t=(R_t,m_t)$: redundancy, placement};
  \node[box, left=of a] (l) {coded burst link, local or edge inference; reward $r_t$};
  \draw[arr] (w)--(q); \draw[arr] (q)--(x); \draw[arr] (x)--(c); \draw[arr] (c)--(a); \draw[arr] (a)--(l);
  \draw[arr, dashed] (l.south) -- ++(0,-4.5mm) -| (c.south);
  \node[font=\scriptsize, anchor=north] at ($(a.south)+(0,-4.5mm)$) {reward feedback};
\end{tikzpicture}
\caption{The PIR loop. Signal quality and cross-modal agreement give $\PIV_t$, which joins channel, energy and compute state in $X_t$; the controller picks redundancy $R_t$ and placement $m_t$, and the realized reward updates it.}
\label{fig:loop}
\end{figure}

\textbf{State.} At window $t$ the controller sees $X_t=[S_m,S_w,S_e,S_c]$. The medical block holds per-modality SQI, presence flags and a scalar value proxy
\begin{equation}
\PIV_t=\operatorname{clip}\!\bigl(w_q\,\mathrm{SQI}_{\mathrm{comb},t}-w_a\,\delta_t,\;0,\;1\bigr),\qquad w_q=0.75,\; w_a=0.25 .
\end{equation}
$\mathrm{SQI}_{\mathrm{comb}}$ is a presence-weighted mean of the per-modality indices (an absent modality is left out of the average, not scored as zero), and $\delta_t$ is the normalized disagreement between ECG- and PPG-derived heart rate, forced to zero unless both are present. The weights are fixed design constants from the signal-processing front end of our earlier work~\cite{kamalakannan2026cardiofusion}, not tuned here. They give $\PIV_t\le w_q=0.75$, which is why clean synthetic windows average 0.74. The scale matters because PIV is traded against energy in the reward; it is neither a probability nor a clinical score.

The wireless block $S_w=[\mathrm{SNR}_t,p_{\mathrm{er},t}]$ follows a two-state Gilbert--Elliott chain~\cite{gilbert1960,elliott1963} with $p_{gb}=0.03$ and $p_{bg}=0.08$, which puts the link in its bad state about 27\% of the time. BPSK over AWGN gives $\mathrm{BER}=Q(\sqrt{2\,\mathrm{SNR}_{\mathrm{lin}}})$, and an $L$-bit payload is erased with $p_{\mathrm{er}}=1-(1-\mathrm{BER})^{L}$. The energy block tracks the battery and a per-window expenditure $E=e_{tx}n_{tx}+e_{rx}n_{rx}+e_{\mathrm{comp}}n_{\mathrm{ops}}$ plus a constant idle draw; the compute block holds on-device and edge inference latency plus transmission latency.

\textbf{Actions and delivery.} The generation size is fixed at $K=32$. An action is a pair $(R,m)$ with $R\in\{2,8,20\}$ redundant coded packets and placement $m\in\{\text{local},\text{edge}\}$, six arms in all. A window is decoded if at least $K$ of the $K+R$ packets survive,
\begin{equation}
\Pdec(K,R,p)=\sum_{k=K}^{K+R}\binom{K+R}{k}(1-p)^{k}\,p^{K+R-k},
\end{equation}
which treats random linear network coding~\cite{ho2006} as an ideal erasure code and ignores the small chance of linear dependence at a finite field size. All controller results use this analytic expression; no packet-level coder is run end to end. At the mean bad-state SNR of 7~dB a 512-bit payload is erased with probability $\approx0.33$, and Eq.~(2) gives $\Pdec\approx 0.00,\,0.06,\,0.85$ for $R=2,8,20$. In the good state all three are indistinguishable from one, so the whole problem lives in the bad state, which is where the controller earns or wastes energy.

\textbf{Utility, reward, policy.} The utility proxy is $U_{\mathrm{PIR}}=\PIV_t\cdot\Pdec$, the window's value discounted by the chance that it is recovered. The reward is
\begin{equation}
r_t=U_{\mathrm{PIR}}-\lambda_E E_t-\lambda_L\,\sd(L_t-L_{\max})-\lambda_P\,\sd(P_t-P_{\max}),
\end{equation}
with softplus penalties that stay near zero inside the 150~ms deadline $L_{\max}$ and the 100~mW cap $P_{\max}$. PIR-LinUCB~\cite{li2010} keeps a ridge estimate $\hat\theta_a=A_a^{-1}b_a$ per arm and plays $a_t=\arg\max_a\, x_t^{\top}\hat\theta_a+\alpha\sqrt{x_t^{\top}A_a^{-1}x_t}$. The context $x_t$ is built from previously observed telemetry (an EWMA erasure estimate, not the channel draw about to be realized), and each update is a rank-one Sherman--Morrison step.

\section{Experimental Protocol}
\label{sec:prot}

\textbf{Synthetic testbed.} A generator of ECG and PPG windows (125~Hz, 8~s) produces 1,200 sequential windows per seed, 200 in each of six regimes: both clean, ECG degraded, PPG degraded, both degraded, ECG missing, PPG missing. Degradation adds severity-scaled Gaussian noise, baseline wander, band-limited muscle artifact and randomly timed motion bursts. The link uses $\mathrm{SNR}\sim\mathcal N(18,1.5^2)$~dB in the good state and $\mathcal N(7,1^2)$~dB in the bad state with a 64-byte coded payload, and every policy sees the same erasure sequence (common random numbers). A BLE-class radio costs $e_{tx}=2$~mJ and $e_{rx}=1$~mJ per packet. Coding costs 8 GF(256) cycles per operation at 0.2~$\mu$J per operation. A 64~MHz microcontroller (2$\times10^{6}$ cycles per inference) is compared with a 1.4~GHz edge target (5$\times10^{6}$ cycles, plus 12~ms round trip), over a 250~kbps PHY. No constant was fitted to a device; the model is a controlled testbed, not a hardware measurement.

\textbf{Policies.} Six policies are run. PIR-LinUCB ($\alpha=0.3$, ridge $\lambda=1$, a ten-dimensional context of per-modality SQI and presence flags, PIV, normalized SNR, EWMA erasure estimate, link state, battery fraction and a bias term; $\lambda_E=4.0$~J$^{-1}$, $\lambda_L=0.05$~ms$^{-1}$, $\lambda_P=0.02$~mW$^{-1}$, softplus sharpness 0.5) is compared with Heuristic-Rule: battery-threshold placement and the smallest $R$ whose decode probability clears 0.99 at the current erasure estimate. The heuristic does not read PIV. PIR-Aware-Heuristic is the same rule extended to read the window's PIV, so that the feature and the learner can be separated. Three static policies complete the set: Fixed-Rate ($R=8$, local) and the pinned $R=20$ policies Always-Local and Always-Edge. All six were run for 25 seeds, each seed on one identical stream.

\textbf{Statistics.} We report steady-state values (second half of each session) as mean $\pm$ 95\% Student-$t$ interval across seeds. Two inferential comparisons are made, each of PIR-LinUCB against a hand-written rule: Heuristic-Rule, and PIR-Aware-Heuristic, the strongest non-bandit policy on reward and on the energy--error trade. Each uses paired $t$-tests across seeds with 95\% intervals of the difference and Holm correction~\cite{holm1979}, applied separately to each comparison over the four outcomes heart-rate MAE, energy, packet delivery ratio (PDR) and reward. Because 25 seeds do not license a normality assumption, we repeat the test with the exact Wilcoxon signed-rank test and count the seeds in which PIR-LinUCB came out ahead and those in which the two policies tie. Other pairs are not tested. To ask whether the controller uses PIV at all, we also ran PIR-LinUCB with the PIV entry of its context replaced by a constant reference value; both variants are scored on the same reward, which contains PIV in either case. Code and per-seed results: \url{https://github.com/ka-cyber/PIR-Framework}.

\section{Results on the Synthetic Testbed}
\label{sec:res}

\begin{table}[t]
\caption{Six policies, 25 seeds, steady state, mean $\pm$ 95\% CI. HR-MAE in bpm, energy in mJ per window, PDR in \%; bold is best per column. Every policy met the 150~ms deadline on every window.}
\label{tab:main}
\centering
\footnotesize
\setlength{\tabcolsep}{2.5pt}
\begin{tabular}{lcccc}
\toprule
Policy & HR-MAE$\downarrow$ & Energy$\downarrow$ & PDR$\uparrow$ & Reward$\uparrow$\\
\midrule
Fixed-Rate ($R{=}8$) & $6.913 \pm 0.303$ & $95.81\pm0.00$ & $82.37\pm2.08$ & $0.1715\pm0.0137$\\
Always-Local ($R{=}20$) & $\mathbf{5.906}\pm0.227$ & $119.89\pm0.00$ & $\mathbf{88.71}\pm1.34$ & $0.1148\pm0.0088$\\
Always-Edge ($R{=}20$)  & $\mathbf{5.906}\pm0.227$ & $119.89\pm0.00$ & $\mathbf{88.71}\pm1.34$ & $0.1167\pm0.0088$\\
Heuristic-Rule          & $6.409\pm0.341$ & $96.66\pm1.38$ & $85.68\pm1.93$ & $0.2015\pm0.0140$\\
PIR-Aware-Heuristic     & $6.110\pm0.231$ & $96.27\pm1.37$ & $87.51\pm1.31$ & $0.2028\pm0.0140$\\
PIR-LinUCB              & $6.100\pm0.231$ & $\mathbf{92.53}\pm1.42$ & $87.56\pm1.31$ & $\mathbf{0.2057}\pm0.0152$\\
\bottomrule
\end{tabular}
\end{table}

\begin{table}[t]
\caption{PIR-LinUCB minus a hand-written rule ($\Delta$), paired over 25 seeds. $p_t$: paired $t$-test; $p_W$: exact Wilcoxon signed-rank test; Holm-adjusted values (family of four) in parentheses. Last column: seeds where PIR-LinUCB was better (ties in parentheses).}
\label{tab:paired}
\centering
\footnotesize
\setlength{\tabcolsep}{1.1pt}
\begin{tabular}{lccccc}
\toprule
Outcome & $\Delta$ & 95\% CI & $p_t$ (Holm) & $p_W$ (Holm) & Better\\
\midrule
\multicolumn{6}{l}{\emph{versus Heuristic-Rule}}\\
HR-MAE, bpm & $-0.309$ & $[-0.494,-0.124]$ & 0.0021 (0.0042) & 5.2e$-$6 (1.6e$-$5) & 23/25\\
Energy, mJ & $-4.135$ & $[-5.228,-3.042]$ & 4.9e$-$8 (1.9e$-$7) & 4.2e$-$7 (1.7e$-$6) & 24/25\\
PDR, pp & $+1.88$ & $[0.94,2.82]$ & 0.0004 (0.0012) & 4.1e$-$5 (8.2e$-$5) & 22/25\\
Reward & $+0.0041$ & $[0.0001,0.0082]$ & 0.045 (0.045) & 0.0056 (0.0056) & 21/25\\
\midrule
\multicolumn{6}{l}{\emph{versus PIR-Aware-Heuristic}}\\
HR-MAE, bpm & $-0.011$ & $[-0.023,0.002]$ & 0.088 (0.18) & 0.028 (0.052) & 6/25 (19)\\
Energy, mJ & $-3.749$ & $[-4.847,-2.651]$ & 2.8e$-$7 (1.1e$-$6) & 5.2e$-$6 (2.1e$-$5) & 24/25\\
PDR, pp & $+0.053$ & $[0.006,0.101]$ & 0.029 (0.088) & 0.024 (0.052) & 6/25 (19)\\
Reward & $+0.0029$ & $[-0.0011,0.0069]$ & 0.146 (0.18) & 0.017 (0.052) & 19/25\\
\bottomrule
\end{tabular}
\end{table}

\begin{figure}[tbp]
\centering
\includegraphics[width=\textwidth]{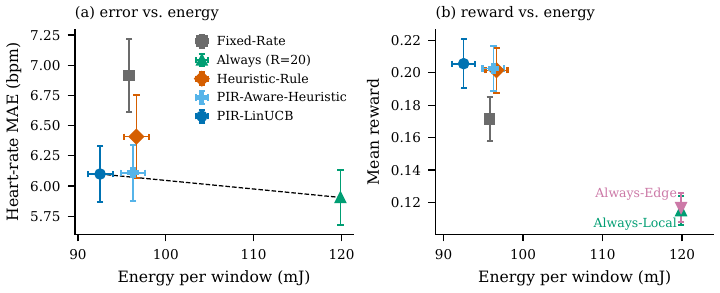}
\caption{Six policies, 25 seeds (means, 95\% CIs). (a)~Heart-rate error versus energy per window (lower left is better); the dashed line joins the two non-dominated policies. Always-Local and Always-Edge coincide in (a). (b)~Mean reward versus energy (upper left is better).}
\label{fig:trade}
\end{figure}

\begin{figure}[tbp]
\centering
\includegraphics[width=\textwidth]{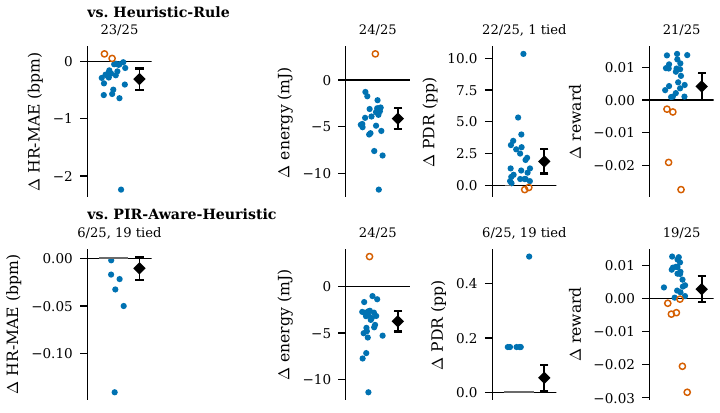}
\caption{Per-seed differences, PIR-LinUCB minus Heuristic-Rule (top) and minus PIR-Aware-Heuristic (bottom). Filled blue dots are seeds where PIR-LinUCB is better, open orange circles where it is worse, gray dashes ties; the diamond is the mean with its 95\% CI. The count above each panel is the number of seeds improved.}
\label{fig:paired}
\end{figure}

\textbf{Against the rule.} PIR-LinUCB came out ahead of Heuristic-Rule on every outcome (Tables~\ref{tab:main} and~\ref{tab:paired}). Energy fell by 4.1~mJ per window (4.3\%), heart-rate error by 0.31~bpm (4.8\%), and packet delivery rose by 1.9 points. These three intervals exclude zero, with Holm-adjusted $p\le0.005$ under the $t$-test and $p\le10^{-4}$ under the rank test. The reward gain, 0.0041 or 2\%, is the thin one: its $t$-test $p$ is 0.045 and its interval only just clears zero, though the rank test is more comfortable (0.0056). The per-seed view (Fig.~\ref{fig:paired}) shows why the tests can disagree: most seeds improve, a few do not, and one contributes a heart-rate gain above 2~bpm. In the energy--error plane (Fig.~\ref{fig:trade}a) PIR-LinUCB beats both Fixed-Rate and Heuristic-Rule on the means, so the saving is not bought with a worse heart-rate estimate.

\textbf{Giving the rule PIV.} The obvious objection is that PIR-LinUCB wins only because it reads PIV and the rule does not. PIR-Aware-Heuristic was run to test that (Table~\ref{tab:main}, lower block of Table~\ref{tab:paired}). Reading PIV closes nearly all of the accuracy and delivery gap. Against Heuristic-Rule, the aware rule lowers heart-rate error by 0.30~bpm (95\% CI $[-0.48,-0.11]$), raises PDR by 1.83 points and reward by 0.0012 (in all 25 seeds), at 0.39~mJ less energy. Against PIR-LinUCB it is indistinguishable on error and delivery: 6.110 against 6.100~bpm and 87.51\% against 87.56\%, and the two policies produce identical values in 19 of 25 seeds. The few seeds that differ drive the rank-test $p$-values, which we do not read as evidence. The reward gap of 0.0029 in favor of PIR-LinUCB has a $t$-test $p$ of 0.15 and a Holm-adjusted rank-test $p$ of 0.052. What does separate them is energy: PIR-LinUCB spends 3.75~mJ (3.9\%) less per window, in 24 of 25 seeds, with Holm-adjusted $p<10^{-4}$ under both tests. The division of credit is therefore specific. In this testbed PIV supplies the gain in accuracy and delivery, and the learned controller supplies an energy saving that a hand-written rule with the same inputs did not find. We have not inspected which actions produce the saving, and we suspect, without having tested it, that it comes from lower redundancy when the link is good.

\textbf{Against static redundancy.} Pinning $R=20$ buys delivery dearly: 88.7\% PDR and 5.91~bpm, but 119.9~mJ per window (30\% above PIR-LinUCB) and the lowest reward, 0.115. PIR-LinUCB gives up 1.2 points of PDR and 0.19~bpm for a 23\% energy saving; this pair was not tested.

\textbf{An inactive action.} From $R=8$ to $R=20$ energy rises by 24.1~mJ for 12 packets, about $e_{tx}$ each, so energy is almost entirely a redundancy decision. Always-Local and Always-Edge match on energy, heart-rate error and PDR; they differ in mean latency (142.1 against 122.3~ms) and by 0.0019 in reward. Placement never binds: every policy meets the deadline, so placement never binds and the problem is in practice a three-arm choice of $R$. We claim nothing about learned placement.

\textbf{Window-specific PIV against a fixed reference.} Giving the controller each window's own PIV raised the common reward by about 0.00156 over the fixed-reference control (95\% CI $[0.000170,\,0.002958]$).  It is tied to the objective: the reward contains PIV by construction, so a feature that predicts PIV-weighted utility is expected to help, and we claim nothing for other objectives. The comparison with PIR-Aware-Heuristic above is what separates the feature from the learner.

\textbf{An earlier run.} An initial five-seed run was too noisy to order the policies on heart-rate error ($6.82\pm1.62$ for PIR-LinUCB against $6.08\pm0.57$~bpm for the heuristic) and is treated as exploratory. In it, steady-state PIV was 0.739 with both modalities clean and 0.543 with both degraded, while error rose from 2.7--3.8 to 13.8--14.4~bpm for every policy. That error, which the controller never sees, is the only outside check the simulation offers on PIV, and it covers two extreme regimes, not a calibration curve.

\section{Does PIV Track Estimation Error on Real Recordings?}
\label{sec:real}

The synthetic study cannot answer this: its generator encodes our own notion of quality. We therefore turned to the Abdominal and Direct Fetal ECG Database~\cite{jezewski2012} (PhysioNet~\cite{goldberger2000}, version~1.0.0, doi:\allowbreak10.13026/C2RP4B). It holds five-minute recordings from five women in labor, each with four abdominal leads sampled at 1~kHz and a simultaneous direct fetal ECG whose R waves were verified by cardiologists. The data are ECG-only, so PPG is absent and the cross-modal term has no counterpart. We use the four leads as redundant channels instead. This is a different sensor setting from the controller study, so it tests the value-proxy idea, not the wireless or compute models, which remain simulated.

\textbf{Method.} Each recording is filtered at 20--100~Hz (zero-phase) and cut into non-overlapping 8~s windows, 37 per recording and 185 in total. For each lead, the fetal rate is taken from the strongest overlap-corrected normalized autocorrelation peak between 105 and 180~bpm, and that peak height serves as the lead's SQI. The window SQI is the mean over leads, the rate estimate is the SQI-weighted consensus of the lead rates, and disagreement is the mean absolute deviation of the lead rates from that consensus. The adapted proxy is
\begin{equation}
\PIV_{\mathrm{ECG}}=\operatorname{clip}\!\bigl(0.75\,\mathrm{SQI}_{\mathrm{comb}}-0.25\,\bar d/140,\;0,\;1\bigr),
\end{equation}
with $\bar d$ the disagreement in bpm. The reference rate and the absolute error come from the direct fetal ECG annotations, which are never an input to PIV. The comparator, called SQI-only below, is $0.75\,\mathrm{SQI}_{\mathrm{comb}}$, that is, the same quantity without the disagreement term; its rank ordering is that of the mean SQI. For each recording we compute the Spearman correlation between the score and the absolute rate error across its 37 windows, then average over recordings with equal weight. We stay within records because pooling would let large differences between mothers masquerade as an effect of window quality. We report no window-level significance tests: windows within a recording are serially dependent, and five recordings are too few for a population-level claim.

\begin{table}[t]
\caption{Within-record Spearman correlation of a window score with absolute fetal heart-rate error (37 windows each); more negative is better.}
\label{tab:real}
\centering
\small
\setlength{\tabcolsep}{5pt}
\begin{tabular}{lcccc}
\toprule
Record & Mean PIV & Mean $|$error$|$ (bpm) & $\rho$, PIV & $\rho$, SQI-only\\
\midrule
r01 & 0.237 & 2.14 & $-0.708$ & $-0.703$\\
r04 & 0.094 & 18.07 & $-0.288$ & $-0.197$\\
r07 & 0.133 & 11.05 & $-0.454$ & $-0.165$\\
r08 & 0.213 & 2.64 & $-0.854$ & $-0.846$\\
r10 & 0.118 & 8.22 & $-0.217$ & $-0.094$\\
\midrule
Mean & -- & -- & $\mathbf{-0.504}$ & $-0.401$\\
\bottomrule
\end{tabular}
\end{table}

\begin{figure}[tbp]
\centering
\includegraphics[width=\textwidth]{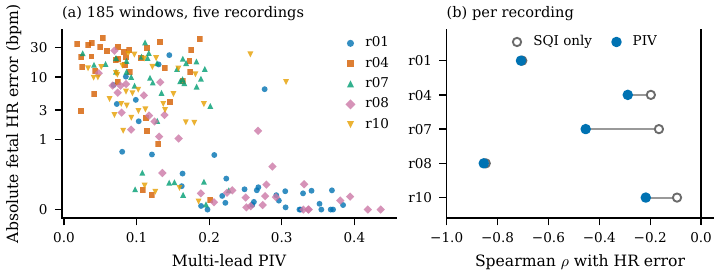}
\caption{Real abdominal fetal ECG (five recordings, 185 windows of 8~s). (a)~Absolute heart-rate error against multi-lead PIV (symmetric-log axis). (b)~Within-record Spearman correlation for PIV (filled) and for the SQI-only comparator (open); the two coincide for r01 and r08.}
\label{fig:real}
\end{figure}

\textbf{Findings.} PIV was inversely related to heart-rate error in all five recordings, with a mean within-record $\rho$ of $-0.504$ against $-0.401$ for SQI-only (Table~\ref{tab:real}, Fig.~\ref{fig:real}). The disagreement term helped in every recording, but unevenly. In r01 and r08 the quality term alone already carried $\rho\approx-0.70$ and $-0.85$, and PIV changed it by less than 0.01. The gain came from r04, r07 and r10 ($\Delta\rho=-0.09$, $-0.29$, $-0.12$), the three recordings where quality alone was weak. We read this as a modest role for lead agreement when the autocorrelation peak is ambiguous, and nothing stronger. Five recordings make even a unanimous direction weak evidence: a two-sided sign test on five of five cannot go below 0.0625.

Grouped by PIV quintile, the median absolute error of the 185 windows falls from 13.80~bpm in the lowest quintile to 6.29, 4.01, 0.78 and 0.13~bpm in the highest. The quintiles pool recordings, so part of that gradient is between-mother variation; the within-record figures are the ones we lean on. The mean within-record correlation stayed negative across four preprocessing settings: $-0.504$ (20--100~Hz, 8~s), $-0.341$ (15--80~Hz), $-0.548$ (25--150~Hz) and $-0.499$ (12~s windows, 125 in all). The narrowest band was weakest, and there r10 changed sign ($\rho=+0.28$), so consistency holds for the mean, not for every recording.

Two cautions apply. First, the quality term and the rate estimate come from the same autocorrelation, so some of the association is built in: a flat or double-peaked autocorrelation yields both a low SQI and a poor rate.  Second, PIV values on this data span only 0.02--0.44, well below the synthetic range, because autocorrelation peaks on abdominal leads are lower than template-correlation scores on clean ECG. The scale of PIV is therefore specific to the sensing front end, and it would have to be recalibrated before being traded against energy in a reward on a new device.

\section{Discussion and Limitations}

The evidence supports a narrow conclusion. A window-level value proxy, channel, energy and compute state can share one state vector and one reward, and a LinUCB controller trained online on that state beats a hand-written rule that ignores PIV modestly but consistently in simulation, on three outcomes by margins that survive two tests. Most of that advantage belongs to PIV: a rule that reads it matches the controller on error and delivery. What remains for learning is a 3.9\% energy saving, and the reward gain over the aware rule is not significant after correction. Under this objective, a controller that sees each window's PIV also earns slightly more than one that sees a constant. On real recordings, multi-lead PIV tracks heart-rate error at least as closely as signal quality, and more closely where quality is weak.

It does not show that the energy saving survives a tuned rule, that the benefit holds under another objective, or that PIV measures clinical value. Both rules are hand-written by us, and one tuned on the same energy price might recover the saving; no oracle bound is reported. The signals in the controller study are synthetic and built from sums of Gaussians, the energy, latency and channel constants are unfitted, decoding is analytic, and only one channel family was tested. The fetal data use four ECG leads rather than ECG and PPG, five recordings, and a rate estimator that shares its input with the quality score. A quality-driven controller may also starve windows where a real arrhythmia distorts the waveform and lowers the score; that risk is untested.

Three extensions follow. Tuning the aware rule against the reward, and logging the actions behind the energy saving, would show whether the learner finds something a rule cannot. A packet-level RLNC coder driven by measured radio traces would replace the analytic decoding model. And a wider range of channels and deadlines, where placement can bind, is needed before learned offloading can be evaluated at all.

\section{Conclusion}

We framed redundancy and placement for a body-worn multimodal node as one online decision made from a window-reliability proxy together with channel, energy and compute state. In 25 paired seeds, a linear contextual bandit lowered energy by 4.3\%, heart-rate error by 0.31~bpm and raised delivery by 1.9 points over a hand-written rule, with a small reward advantage. Most of that advantage came from the PIV input: a rule given PIV matched the bandit on error and delivery, and the bandit's remaining edge is a 3.9\% lower energy draw. Placement stayed inactive, and the PIV ablation is small and tied to the objective. On five recorded fetal ECGs the proxy followed estimation error, and followed it better than signal quality alone in every record, at a weight that five recordings cannot turn into a general claim. The natural next test is a tuned rule against the learner, evaluated on measured radio traces.

\paragraph{Ethics statement.} The fetal ECG recordings are de-identified and distributed through PhysioNet under the Open Data Commons Attribution License v1.0; no new human-subject data were collected, and no clinical claim is made.

\paragraph{Disclosure of Interests.} The authors have no competing interests to declare that are relevant to this article.

\bibliographystyle{splncs04}
\bibliography{refers}

@article{elliott1963,
  author  = {Elliott, E. O.},
  title   = {Estimates of error rates for codes on burst-noise channels},
  journal = {Bell System Technical Journal},
  volume  = {42},
  number  = {5},
  pages   = {1977--1997},
  year    = {1963}
}

@article{gilbert1960,
  author  = {Gilbert, E. N.},
  title   = {Capacity of a burst-noise channel},
  journal = {Bell System Technical Journal},
  volume  = {39},
  number  = {5},
  pages   = {1253--1265},
  year    = {1960}
}

@article{gunduz2023,
  author  = {G{\"u}nd{\"u}z, Deniz and Qin, Zhijin and Aguerri, I{\~n}aki Estella and Dhillon, Harpreet S. and Yang, Zhaohui and Yener, Aylin and Wong, Kai Kit and Chae, Chan-Byoung},
  title   = {Beyond transmitting bits: Context, semantics, and task-oriented communications},
  journal = {IEEE Journal on Selected Areas in Communications},
  volume  = {41},
  number  = {1},
  pages   = {5--41},
  year    = {2023}
}

@article{ho2006,
  author  = {Ho, Tracey and M{\'e}dard, Muriel and Koetter, Ralf and Karger, David R. and Effros, Michelle and Shi, Jun and Leong, Ben},
  title   = {A random linear network coding approach to multicast},
  journal = {IEEE Transactions on Information Theory},
  volume  = {52},
  number  = {10},
  pages   = {4413--4430},
  year    = {2006}
}

@article{kosta2017,
  author  = {Kosta, Antzela and Pappas, Nikolaos and Angelakis, Vangelis},
  title   = {Age of information: A new concept, metric, and tool},
  journal = {Foundations and Trends in Networking},
  volume  = {12},
  number  = {3},
  pages   = {162--259},
  year    = {2017}
}

@inproceedings{li2010,
  author    = {Li, Lihong and Chu, Wei and Langford, John and Schapire, Robert E.},
  title     = {A contextual-bandit approach to personalized news article recommendation},
  booktitle = {Proceedings of the 19th International Conference on World Wide Web (WWW)},
  pages     = {661--670},
  publisher = {ACM},
  year      = {2010},
  doi       = {10.1145/1772690.1772758}
}

@article{liang2024,
  author  = {Liang, Paul Pu and Zadeh, Amir and Morency, Louis-Philippe},
  title   = {Foundations and trends in multimodal machine learning: Principles, challenges, and open questions},
  journal = {ACM Computing Surveys},
  volume  = {56},
  number  = {10},
  pages   = {1--42},
  year    = {2024},
  doi     = {10.1145/3656580}
}

@inproceedings{ma2022,
  author    = {Ma, Mengmeng and Ren, Jian and Zhao, Long and Testuggine, Davide and Peng, Xi},
  title     = {Are multimodal transformers robust to missing modality?},
  booktitle = {Proceedings of the IEEE/CVF Conference on Computer Vision and Pattern Recognition (CVPR)},
  pages     = {18177--18186},
  year      = {2022}
}

@article{orphanidou2015,
  author  = {Orphanidou, Christina and Bonnici, Timothy and Charlton, Peter and Clifton, David and Vallance, David and Tarassenko, Lionel},
  title   = {Signal-quality indices for the electrocardiogram and photoplethysmogram: Derivation and applications to wireless monitoring},
  journal = {IEEE Journal of Biomedical and Health Informatics},
  volume  = {19},
  number  = {3},
  pages   = {832--838},
  year    = {2015},
  doi     = {10.1109/JBHI.2014.2338351}
}

@inproceedings{zhang2023,
  author    = {Zhang, Qingyang and Wu, Haitao and Zhang, Changqing and Hu, Qinghua and Fu, Huazhu and Zhou, Joey Tianyi and Peng, Xi},
  title     = {Provable dynamic fusion for low-quality multimodal data},
  booktitle = {Proceedings of the 40th International Conference on Machine Learning},
  series    = {Proceedings of Machine Learning Research},
  volume    = {202},
  pages     = {41753--41769},
  year      = {2023}
}

@inproceedings{chu2011,
  author    = {Chu, Wei and Li, Lihong and Reyzin, Lev and Schapire, Robert E.},
  title     = {Contextual bandits with linear payoff functions},
  booktitle = {Proceedings of the 14th International Conference on Artificial Intelligence and Statistics (AISTATS)},
  series    = {Proceedings of Machine Learning Research},
  volume    = {15},
  pages     = {208--214},
  year      = {2011}
}

@inproceedings{abbasi2011,
  author    = {Abbasi-Yadkori, Yasin and P{\'a}l, D{\'a}vid and Szepesv{\'a}ri, Csaba},
  title     = {Improved algorithms for linear stochastic bandits},
  booktitle = {Advances in Neural Information Processing Systems 24},
  pages     = {2312--2320},
  year      = {2011}
}

@inproceedings{agrawal2013,
  author    = {Agrawal, Shipra and Goyal, Navin},
  title     = {Thompson sampling for contextual bandits with linear payoffs},
  booktitle = {Proceedings of the 30th International Conference on Machine Learning},
  series    = {Proceedings of Machine Learning Research},
  volume    = {28},
  pages     = {127--135},
  year      = {2013}
}

@inproceedings{zhou2020,
  author    = {Zhou, Dongruo and Li, Lihong and Gu, Quanquan},
  title     = {Neural contextual bandits with {UCB}-based exploration},
  booktitle = {Proceedings of the 37th International Conference on Machine Learning},
  series    = {Proceedings of Machine Learning Research},
  volume    = {119},
  pages     = {11492--11502},
  year      = {2020}
}

@article{shao2022,
  author  = {Shao, Jiawei and Mao, Yuyi and Zhang, Jun},
  title   = {Learning task-oriented communication for edge inference: An information bottleneck approach},
  journal = {IEEE Journal on Selected Areas in Communications},
  volume  = {40},
  number  = {1},
  pages   = {197--211},
  year    = {2022},
  doi     = {10.1109/JSAC.2021.3126087}
}

@article{holm1979,
  author  = {Holm, Sture},
  title   = {A simple sequentially rejective multiple test procedure},
  journal = {Scandinavian Journal of Statistics},
  volume  = {6},
  number  = {2},
  pages   = {65--70},
  year    = {1979}
}

@article{goldberger2000,
  author  = {Goldberger, Ary L. and Amaral, Luis A. N. and Glass, Leon and Hausdorff, Jeffrey M. and Ivanov, Plamen Ch. and Mark, Roger G. and Mietus, Joseph E. and Moody, George B. and Peng, Chung-Kang and Stanley, H. Eugene},
  title   = {{PhysioBank}, {PhysioToolkit}, and {PhysioNet}: Components of a new research resource for complex physiologic signals},
  journal = {Circulation},
  volume  = {101},
  number  = {23},
  pages   = {e215--e220},
  year    = {2000}
}

@article{jezewski2012,
  author  = {Jezewski, Janusz and Matonia, Adam and Kupka, Tomasz and Roj, Dawid and Czabanski, Robert},
  title   = {Determination of fetal heart rate from abdominal signals: Evaluation of beat-to-beat accuracy in relation to the direct fetal electrocardiogram},
  journal = {Biomedizinische Technik / Biomedical Engineering},
  volume  = {57},
  number  = {5},
  pages   = {383--394},
  year    = {2012},
  doi     = {10.1515/bmt-2011-0130}
}

@misc{kamalakannan2026cardiofusion,
  author = {Kamalakannan, Navaneetha Krishnan and Kamalakannan, Janakiraman},
  title  = {{CardioFusion-AI}: Robust {ECG--PPG} fusion for multimodal physiological monitoring under signal degradation},
  year   = {2026},
  note   = {arXiv:2608.26000}
}

@misc{kamalakannan2026apc,
  author = {Kamalakannan, Navaneetha Krishnan and Velmurugan, Harinisri},
  title  = {Adaptive peer clustering with hierarchical random linear network coding for resilient decentralized wireless networks},
  year   = {2026},
  note   = {arXiv:2608.26040}
}

\end{document}